# Gamma-spectrometric determination of $^{232}U$ in uranium-bearing materials

Jozsef Zsigrai[1], Tam Cong Nguyen[2], Andrey Berlizov[1*+]

[1]European Commission, Joint Research Centre (JRC), Institute for Transuranium Elements (ITU), 76125 Karlsruhe, P.O. Box 2340, Germany
[2]Centre for Energy Research of the Hungarian Academy of Sciences (EK), 1525 Budapest 114, P.O. Box 49, Hungary

**Abstract**

The $^{232}U$ content of various uranium-bearing items was measured using low-background gamma spectrometry. The method is independent of the measurement geometry, sample form and chemical composition. Since $^{232}U$ is an artificially produced isotope, it carries information about previous irradiation of the material, which is relevant for nuclear forensics, nuclear safeguards and for nuclear reactor operations. A correlation between the $^{232}U$ content and $^{235}U$ enrichment of the investigated samples has been established, which is consistent with theoretical predictions. It is also shown how the correlation of the mass ratio $^{232}U/^{235}U$ vs. $^{235}U$ content can be used to distinguish materials contaminated with reprocessed uranium from materials made of reprocessed uranium.

Keywords: $^{232}U$, gamma spectrometry, reprocessed uranium, nuclear safeguards, nuclear forensics

## 1. Introduction

In this work a non-destructive method is presented for measuring the $^{232}U$ content of uranium-bearing items by gamma spectrometry. The original aim was to extend the nuclear forensics toolbox helping to trace the origin and history of illicit nuclear material. However, the method can also be applied in other fields where the knowledge of the $^{232}U$ content is relevant, such as nuclear reactor operation, nuclear safeguards and nuclear arms control.

For nuclear forensics it is important that $^{232}U$ typically does not occur in natural uranium in measurable quantities, but is formed during the irradiation of uranium or thorium in a nuclear reactor. Therefore, if $^{232}U$ is found then it carries information about the history of the material in which it was detected. If $^{232}U$ is present, then it means that the sample contains some irradiated material (e.g. reprocessed uranium) or is contaminated with such [1]. This information helps to trace the origin of illicit nuclear material [2].

For nuclear-reactor operators the presence of $^{232}U$ in uranium fuel is relevant because it implicitly implies the presence of $^{236}U$, which is a neutron absorber and influences the operation of a nuclear reactor (see, e.g. page 11 in [3]). While the direct measurement of $^{236}U$ in reactor fuel is only possible by destructive methods, $^{232}U$ in reactor fuel can be measured non-destructively by gamma spectrometry [4].

The detection of the isotope $^{232}U$ had also been proposed for confirming the presence and distribution of highly enriched uranium (HEU) in nuclear weapons [5]. In [6] it has been argued that the presence of $^{232}U$ can be an unclassified attribute of HEU for nuclear arms

*Present address: International Atomic Energy Agency, Vienna International Centre, PO Box 100, 1400 Vienna, Austria
+Former address: Institute for Nuclear Research, Prospekt Nauky, 47, 03680, Kyiv, Ukraine

control. Furthermore, ideas on using small amounts of $^{232}U$ added to uranium have been proposed in [7] and [8] to help prevent the proliferation of nuclear weapons.

Two standards (see [9]), ASTM C 996 [10] and ASTM C 787 [11] define the limits on the $^{232}U$ and $^{236}U$ contents of the feed to an enrichment process and of uranium enriched to less than 5% $^{235}U$. For natural U the limit is defined relative to total U, while for enriched U it is defined relative to $^{235}U$, as follows:

- "Commercial natural uranium" (CNU) [10]:
  - $^{232}U$ content < 1 x $10^{-11}$ g/g U and $^{236}U$ content < 2 x 10-5 g/g U
- "Enriched commercial grade uranium" (ECGU) [11]:
  - $^{232}U$ content < 2 x $10^{-9}$ g/g $^{235}U$ and $^{236}U$ content < 5 x $10^{-3}$ g/g $^{235}U$.

Upon enrichment to 5% $^{235}U$ "Commercial natural uranium" (CNU) will become "Enriched commercial grade uranium" (ECGU) which satisfies the specifications of ASTM-C996. Lower $^{232}U$ and $^{236}U$ contents are considered to be due to trace contamination by irradiated uranium and, from the point of view of transport, storage and handling, the material is treated as unirradiated uranium of natural origin.

The method described in this paper can show whether a sample satisfies the criteria of ASTM C787 and ASTM C996 for CNU and ECGU. To support nuclear-forensic investigations, even minor traces of reprocessed uranium in ECGU can be detected proving if a sample originates from a facility which handles reprocessed uranium.

The method is applicable to a wide range of samples, from the smallest ones containing less than 1 g of uranium, to complete nuclear-reactor fuel assemblies. The measurable $^{232}U$ content varies in a range of 4 orders of magnitude. If the sample is homogenous, then the $^{232}U$ content obtained by the described method is independent of the measurement geometry, sample form and chemical composition.

## 2. Theoretical prediction of the $^{232}U$ content of Uranium

The artificial nuclide $^{232}U$ can form in a variety of nuclear reaction chains. The most important were given, for example, in [6], [12], [13], [14] and [15]. These reactions occur, e.g., during the irradiation of uranium or thorium fuel in a nuclear reactor.

To estimate the $^{232}U$ content of spent nuclear fuel we used the webKORIGEN depletion calculation engine available within Nucleonica [16]. We calculated the approximate $^{232}U$ content remaining in spent reactor fuel 6 years after the end of irradiation for fuel of 4% initial $^{235}U$ enrichment[1]. For spent pressurized water reactor fuel for a range of burn-ups between 15 to 60 MWd/kgU we calculated that there is $3.82\times10^{-8}$ to $5.82\times10^{-7}$ mass % of $^{232}U$ relative to total U. We also estimated that the $^{232}U$ content relative to the remaining $^{235}U$ is in the range from $1.48\times10^{-8}$ to $1.40\times10^{-6}$ g/g $^{235}U$. These values are in accordance with previous estimates reported in [14] and [15].

In [13] the $^{232}U$ content of uranium was estimated from burn-up calculations and from simple mathematical models of the enrichment cascades. It was concluded in [13] that depleted uranium contains 1600 to 8000 times less $^{232}U$ than HEU. Furthermore, it has been estimated in [13] that cascade enrichment increases the $^{232}U$ concentration by a factor of 200 to 1000.

[1] The webKORIGEN settings used for the calculations were the following: Mode of calculation: Reactor irradiation and decay; Reactor type: PWR; Fuel: uranium oxide with 4.0% enrichment; Cross section library: "PWR UOX 4.0% U235 60 MWd/kgHM"; Length of cycle: 1y; Number of cycles: 2 for 15 MWd/kgU and 5 for 50-60 MWd/kgU; Load factor: 80.0%; Fuel decay time after discharge: 6 y, Heavy metal mass: 20 t.

When uranium from reprocessed spent fuel is used to make new fuel for nuclear reactors, it is usually blended with other uranium materials to adjust the $^{235}U$ enrichment of the product to a specified value. Therefore, the final $^{232}U$ content of the product is less than that of the spent fuel.

# 3. Instruments and materials

For the studies presented in this paper spectra were taken with 4 different HPGe detectors at 4 different locations (see Table 1).

**Table 1.** Gamma spectrometers used in this work

| Detector short name | Location | Manufacturer | Measured FWHM at 1332 keV (kev) | Declared efficiency (%)* |
|---|---|---|---|---|
| EK1 | Budapest, Hungary | PGT | 2.05 | 34 |
| EK2 | Budapest, Hungary | Canberra | 1.82 | 35 |
| INR | Kyiv, Ukraine | Canberra | 1.78 | 63 |
| ITU1 | Karlsruhe, Germany | Canberra | 1.86 | 54 |
| ITU2 | Karlsruhe, Germany | Ortec | 1.78 | 52 |

*In the standard definition, at 1332 keV and 25 cm source-detector distance, relative to a 3" x 3" NaI(Tl) detector.

The majority of the spectra were taken by a low-background HPGe detector (EK1) located at the Department for Nuclear Security of the Centre for Energy Research in Budapest, Hungary. Spectra of research-reactor fuel rods were taken with a detector (EK2) on site of the research reactor of the Centre for Energy Research. Spectra of certified reference materials were also taken by a detector at the Institute of Nuclear Research in Kyev, Ukraine (INR). Finally, various spectra were taken at the Institute of Transuranium Elements of the Joint Research Centre of the European Commission in Karlsruhe, Germany (detectors ITU1 and ITU2). Table 1 summarizes the detectors used.

The items investigated in this work are listed in Table 5 in the Appendix, together with some basic information about them. The samples included, among others, certified reference materials, seized fuel pellets, research-reactor fuel rods and U metal, spanning an enrichment range from depleted to highly enriched uranium.

# 4. The method for measuring the $^{232}U$ content of uranium by gamma spectrometry

## *4.1. General description of the method*

In most cases the gamma radiation coming directly from $^{232}U$ cannot be detected by gamma spectrometry because the $^{232}U$ gamma peaks are masked by the Compton background of the peaks from the major uranium isotopes. However, the daughter products of $^{232}U$, in particular $^{212}Pb$, $^{212}Bi$, and $^{208}Tl$ emit strong gamma radiation detectable by gamma spectrometry. All these three isotopes are short-lived and they are in equilibrium with $^{228}Th$.

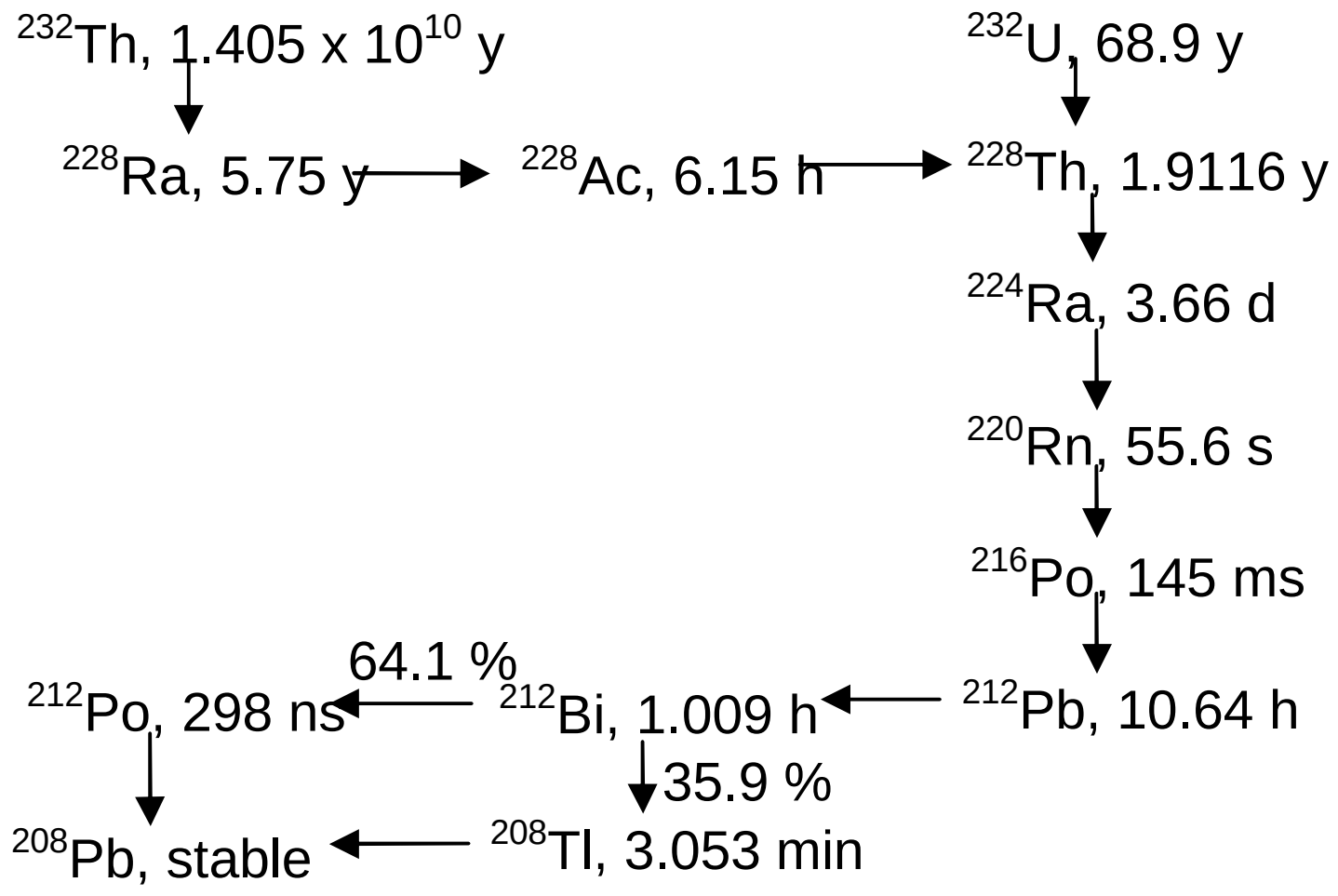

**Figure 1.** The decay chain of $^{232}U$ and $^{232}Th$

These isotopes, however, are also present in the decay chain of $^{232}Th$ (see Figure 1), and the presence of the gamma-emitting nuclides $^{212}Pb$, $^{212}Bi$, and $^{208}Tl$ might be also due to the presence of $^{232}Th$ in the sample. The two decay chains merge at $^{228}Th$. Therefore, the activity of $^{228}Th$, as well as of its short-lived gamma-emitting daughters $^{212}Pb$, $^{212}Bi$ and $^{208}Tl$, can be given as the sum of two terms: one of them accounting for the build-up from $^{232}Th$ and another accounting for the build-up from $^{232}U$. This is reflected in the following equation, which can be obtained using the Bateman solution to the equations of radioactive decay [17]:

$$
\begin{aligned}
\frac{A_{Tl208}}{p} &= A_{Bi212} = A_{Pb212} = A_{Th228} = \\
&= A_{Th232}\left[1 + \frac{\lambda_{Th228}\exp(-\lambda_{\mathrm{Ra}\,228}t) - \lambda_{Ra228}\exp(-\lambda_{Th228}t)}{\lambda_{Ra228} - \lambda_{Th228}}\right] + \\
&+ A_{U232}\left[\frac{1-\exp((\lambda_{U232} - \lambda_{Th228})t)}{1-\lambda_{U232}/\lambda_{Th228}}\right]
\end{aligned}
\qquad (1)
$$

where $p$ = 0.359 [16] is the decay branching probability of the decay of $^{212}Bi$ to $^{208}Tl$ (see Figure 1), $\lambda_{Th228}$, $\lambda_{Ra228}$ and $\lambda_{Th232}$ are the respective decay constants, while $A_{Tl208}$, $A_{Bi212}$, $A_{Pb212}$, $A_{Th228}$, $A_{Th232}$ and $A_{U232}$ are the corresponding activities at the time of the measurement. Note that equation (1) was obtained taking into account that $\lambda_{Th228} >> \lambda_{Th232}$ and $\lambda_{Ra228} >> \lambda_{Th232}$.

Observing the decay scheme of $^{232}U$ and $^{232}Th$ (Figure 1) and the gamma energies emitted by their daughters (Table 2), it can be seen that the activity of $^{232}Th$ can be calculated from the activity of the short lived-isotope $^{228}Ac$, which, in turn, can be determined from the gamma peaks at 911.316 and 969.171 keV. Using the law of radioactive decay and assuming secular equilibrium between the short-lived daughters and their parent isotope, one obtains

$$A_{Ac228} = A_{Ra228} = A_{Th232}\left(1-\exp(-\lambda_{Ra228}t)\right) \qquad (2)$$

where $A_{Ac228}$, $A_{Ra228}$ and $A_{Th232}$ denote the corresponding activities, $\lambda_{Ra228}$ is the decay constant of $^{228}$Ra, while $t$ is the age of the sample (under sample age we mean the time passed since the daughter products were separated from the parent isotopes).

Using the measured activities of $^{228}$Ac, $^{212}$Bi and $^{208}$Tl, the measured or estimated age of the sample and the known decay constants (half-lives), the activities of $^{232}$U and $^{232}$Th are obtained by solving Eqs. (1) and (2) for $A_{U232}$ and $A_{Th232}$ [1][18].

Subtracting the background count rate for each peak was extremely important for most of the measured samples, as the background count rate was of the same order of magnitude, as the count rate coming from the samples, despite using a low-background measurement setup.

### *4.2. Using relative efficiency curve to calculate $^{232}$U content*

In this work the activities of $^{228}$Ac, $^{212}$Bi and $^{208}$Tl were measured relative to $^{238}$U, using relative efficiency calibration. Consequently, the $^{232}$U activity was also obtained relative to $^{238}$U. Then, using the known $^{238}$U isotopic fraction of the samples, the $^{232}$U ratio relative to total uranium was calculated.

The energies and emission probabilities of the gamma peaks used in this work are listed in Table 2. Note that the 609 keV peak of $^{214}$Bi is not directly used for determining the $^{232}$U content, but it is relevant for determining the age of those samples for which this information is not available from the certificate or from destructive analysis.

**Table 2.** The energies and emission probabilities of the gamma peaks used in this work [16]. The emission probabilities of $^{234}$Pa and $^{234m}$Pa are normalized per decay of $^{238}$U.

| Energy (keV) | Emission probability (%) | Emitter |
|---|---|---|
| $^{238}$U group | | |
| 569.15 | 0.0154 ± 0.0013 | $^{234}$Pa |
| 766.36 | 0.3193 ± 0.0033 | $^{234m}$Pa |
| 1001.02 | 0.8350 ± 0.0000 | $^{234m}$Pa |
| 1193.77 | 0.01311 ± 0.00042 | $^{234m}$Pa |
| 1510.10 | 0.01303 ± 0.00025 | $^{234m}$Pa |
| 1737.80 | 0.02121 ± 0.00017 | $^{234m}$Pa |
| 1831.70 | 0.01728 ± 0.00017 | $^{234m}$Pa |
| $^{228}$Th group | | |
| 583.191 | 85.1 ± 0.6 | $^{208}$Tl |
| 727.33 | 6.74 ± 0.12 | $^{212}$Bi |
| 860.566 | 12.52 ± 0.12 | $^{208}$Tl |
| 2614.55 | 99.83 ± 0.17 | $^{208}$Tl |
| $^{228}$Ac group | | |
| 911.316 | 29 ± 0 | $^{228}$Ac |
| 969.161 | 17.45 ± 1.74 | $^{228}$Ac |
| $^{226}$Ra group | | |
| 609.318 | 46.89 ± 4.00 | $^{214}$Bi |

The background-corrected count rates, $C_{\gamma,N}$, at the energies listed in Table 2, coming from nuclide "N", were divided by the corresponding emission probabilities, $I_{\gamma,N}$. This way one obtains the normalized count rates, $K_{\gamma,N}$, defined as:

$$K_{\gamma,N} = \frac{C_{\gamma,N}}{I_{\gamma,N}} \quad . \qquad (3)$$

To get the relative efficiency function the values of $\ln(K_{\gamma,N})$ for the 7 peaks of $^{238}U$ daughters $^{234}Pa$ and $^{234m}Pa$ listed in Table 2 were plotted as a function of the natural logarithm of radiation energy. A linear function was fitted to the data points on the ln-ln scale using weighted least squares fitting, so that the data points with larger statistical uncertainties have less influence on the fit. The straight line proved to fit very well to the data points, i.e. the fit follows very accurately the shape of the detector efficiency in the investigated energy region between 569 and 1831 keV (see Figure 2).

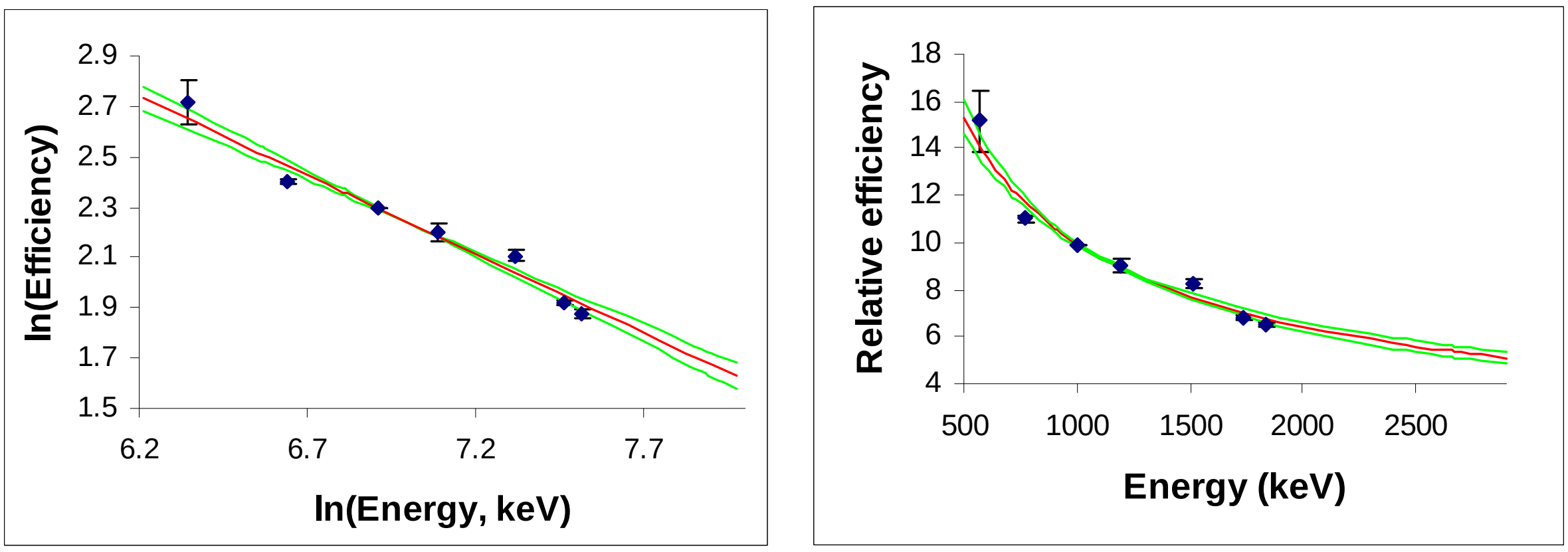

a) b)

**Figure 2.** Relative efficiency function of the "EK1" detector on logarithmic (a) and linear scale (b) constructed using the spectrum of LEU pellets (sample "642" in Table 5). The 67 % confidence bands are also shown.

This curve was also extrapolated to the energy of the strongest gamma line of $^{208}Tl$, to 2614 keV. The uncertainty of the extrapolated efficiency, however, is obviously larger than in the region between the efficiency data points, as shown by the confidence bands in Figure 2.

Let us denote the value of the fitted relative efficiency function constructed from the peaks of $^{238}U$ daughters at energy $\gamma$ as $f_{238}(\gamma)$. Then the ratio, $A_N/A_{238}$, of the activity of the nuclide "N" to the activity of $^{238}U$ can be given as

$$\frac{A_N}{A_{238}} = \frac{K_{\gamma,\mathrm{N}}}{f_{238}(\gamma)} \qquad (4),$$

where $K_{\gamma,\mathrm{N}}$ is the normalized count rate defined by equation (3). To get the final value for the activity of a nuclide the weighted average of the values calculated at different energies was used. For example, the activity of $^{228}Th$ (relative to $^{238}U$) used in equation (1) was calculated as the weighted average of the 4 values obtained from the 4 gamma lines of its daughters listed in Table 2.

For homogeneous samples equation (4) ensures that the calculated activity ratio is independent of the measurement geometry and chemical composition of the sample. Namely, if the count rate $K_{\gamma,\mathrm{N}}$ at a specific energy would change due to a change of geometry or chemical composition, then the relative efficiency $f_{238}(\gamma)$ would change by the same factor. Therefore, the ratio will be unchanged.

In order to account for the possible $^{232}Th$ content of the sample, we monitored the count rate of the gamma peaks of $^{228}Ac$. The activity ratio of $^{228}Ac$ to $^{238}U$ was obtained as the weighted average of the values calculated using the count rates of the gamma lines at 911.204 and 968.971 keV in equation (4). Then the ratio of the activities of $^{232}Th$ and $^{238}U$ was determined from formula (2) taking into account the age of the material.

The $^{232}U$ to $^{238}U$ ratio was calculated from equation (1), using the measured ratios of $^{228}Th$ and $^{232}Th$ to $^{238}U$ and the known age of the material.

Finally, the $^{232}U$ content of uranium was obtained by multiplying the measured $^{232}U$ to $^{238}U$ mass ratio by the known $^{238}U$ mass fraction of the investigated items. The $^{238}U$ fraction was measured either by gamma spectrometry or by mass spectrometry, while for the certified reference materials it was available from the certificate.

### *4.3. Uncertainty calculation*

The uncertainties where calculated by propagating the uncertainties of all the quantities entering into the equation for calculating the $^{232}U$ content. These were the following: the counting uncertainties of the peak areas in the spectra of the samples and of the background, the uncertainties of the emission probabilities, the uncertainty of sample age, the uncertainty of calculating the contribution of $^{232}Th$, and the statistical uncertainty of the fit to the relative efficiency function. The width of the 67 % confidence bands was taken as the uncertainty of the fitted relative efficiency function at a given energy. All uncertainties are given with a coverage factor of $k=1$, i.e., at the 67 % confidence level ("1 sigma").

## 5. Results

### *5.1. The measured $^{232}U$ content of the investigated items*

The results of the measurement of the $^{232}U$ content are shown on Figure 4 and Figure 5 and they are also given in detail in Table 6 in the Appendix. In order to compare the $^{232}U$ content of materials with different ages, we also calculated their $^{232}U$ content at the time of their production, using the available age values (declared, measured or estimated). This $^{232}U$ content is also given in Table 6 and its dependence on $^{235}U$ enrichment is given in Figure 3 (the values for the items made of natural uranium are not shown in the figure, because they were below detection limit).

It can be seen from Figure 3 that, in general, the measured $^{232}U$ content of the samples increases with $^{235}U$ enrichment. However, for a certain number of items the $^{232}U$ content is about two orders of magnitude higher than for other items containing uranium of similar $^{235}U$ enrichment.

In order to compare the $^{232}U$ contents of the investigated samples to the limits set in the standards ASTM C 996 [10] and ASTM C 787 [11], Figure 4 shows the $^{232}U/^{235}U$ ratio for all the measured samples. The horizontal line in Figure 4 represents the limit of $2x10^{-9}$ g/g $^{235}U$ for "Enriched commercial grade uranium", defined in ASTM C 996. Furthermore, it can be seen in Table 6 that the $^{232}U$ content of samples of natural isotopic composition is less than the limit of $10^{-11}$ g/g U for "Commercial natural uranium" defined in ASTM C 787.

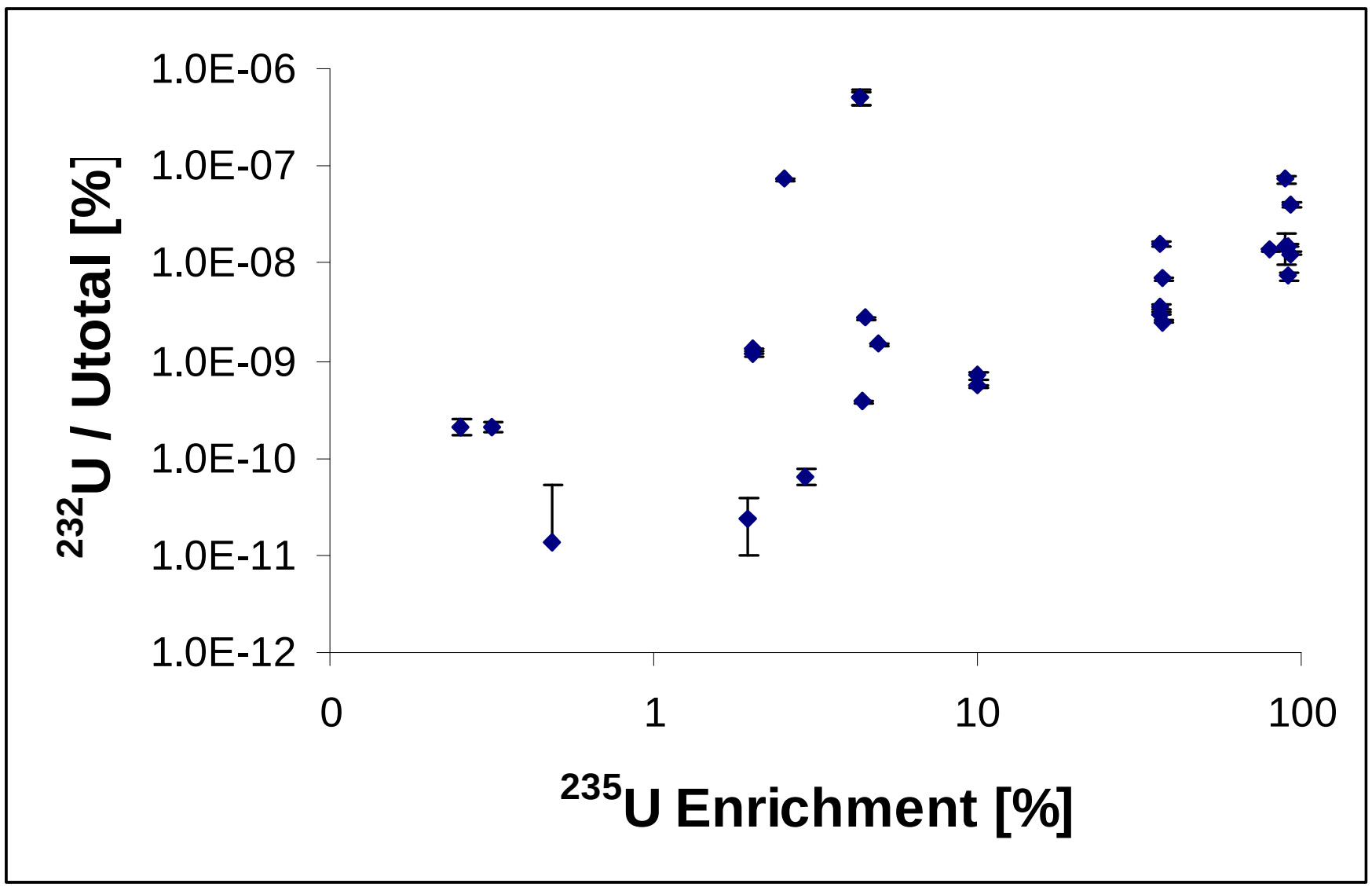

**Figure 3.** The dependence of the measured $^{232}U$ content (extrapolated to the date of production) on the $^{235}U$ content. Since $^{232}U$ can be present only if the investigated item contains recycled uranium, all samples on this plot contained at least traces of recycled uranium. The plot also confirms the conjecture from [13] that the $^{232}U$ content increases with $^{235}U$ enrichment.

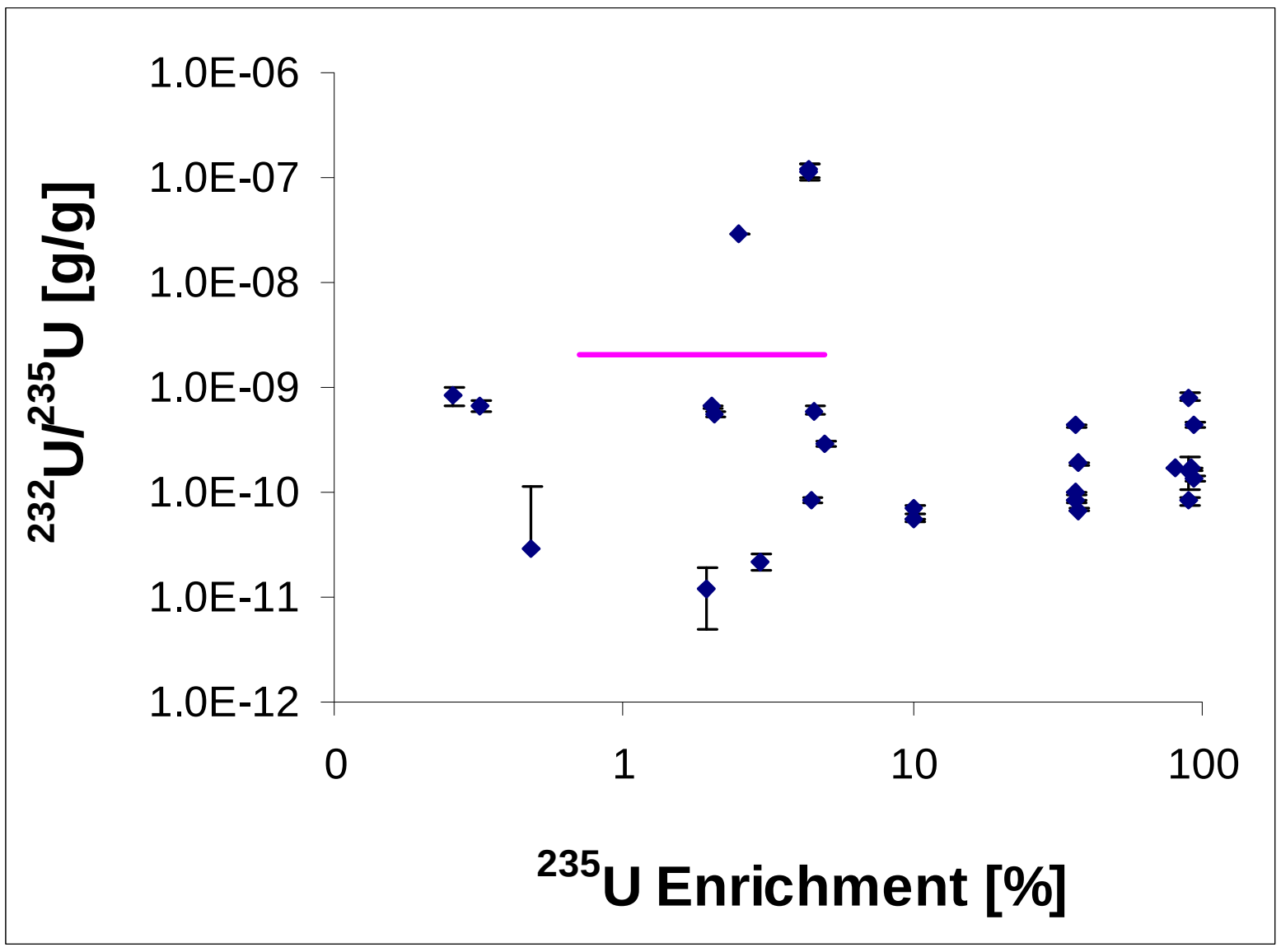

**Figure 4.** The $^{232}U/^{235}U$ ratio for the investigated samples as a function of $^{235}U$ enrichment. For most samples, the $^{232}U/^{235}U$ mass ratio was below < $2x10^{-9}$ g/g $^{235}U$, which is the limit for “Enriched commercial grade uranium” (horizontal line on the graph). The exceptions above the line can be considered as "made of reprocessed uranium" (Romans1 and Romans2 overlap). This means that reprocessed uranium was added during their production. For all the rest $^{232}U$ is probably present only due to cross-contamination in the enrichment cascade.

For some of the investigated materials the $^{236}U$ content was also known, either from a certificate, or from parallel mass-spectrometric measurements. The items with exceptionally high $^{232}U$ content had also exceptionally high $^{236}U$ content. Analogously to $^{232}U$, $^{236}U$ is also an indicator of recycled uranium, confirming that at least some recycled uranium was used in the production of the samples of high $^{232}U$ content.

*5.2. Reproducibility of the results*

The spectra of some of the materials were taken in parallel at different locations, with different detectors, to investigate the reproducibility of the results obtained with the described method. From these spectra the $^{232}U$ content was calculated using the above described procedure.

The spectra of the "CBNM" set of certified reference materials [19] (see Table 5 in the Appendix) were taken with the detectors "EK1", "INR" and "ITU1" from Table 1. The measured $^{232}U$ contents agree very well, within the "1 sigma" measurement uncertainty, as it can be seen in Figure 5.

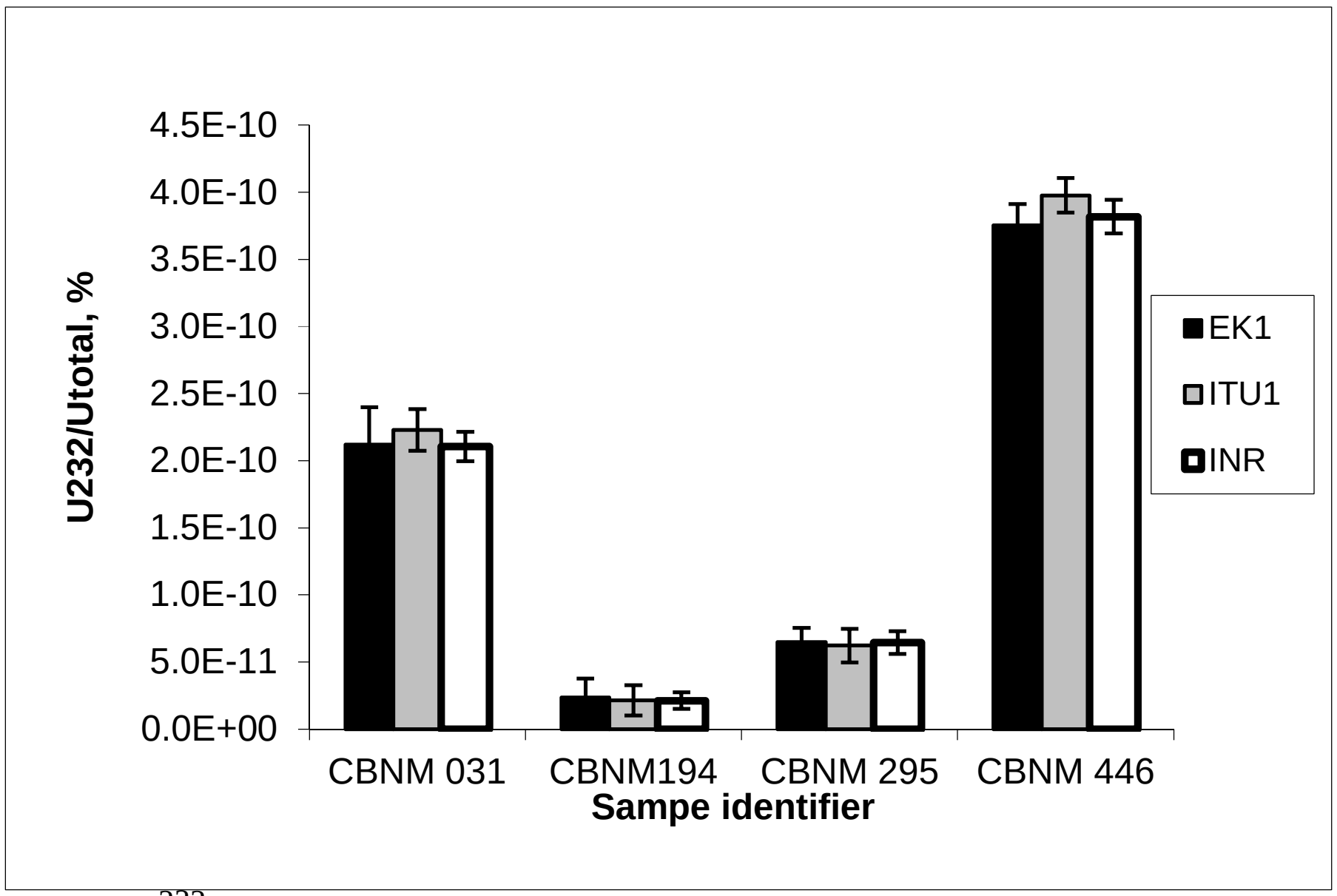

**Figure 5.** The $^{232}U$ content of the CBNM standards measured with three different detectors, on three different locations, extrapolated to the time of production of the standards.

**Table 3.** Comparative measurement of various materials at different locations

| Sample | $^{232}U$ content extrapolated to the time of production, measured at different locations | |
|---|---|---|
| | EK1 | ITU1 |
| HU642 | $(7.27 \pm 0.12)\times10^{-8}$ | $(7.26 \pm 0.11)\times10^{-8}$ |
| HU643 | $(2.13 \pm 0.41)\times10^{-10}$ | $(1.73 \pm 0.27)\times10^{-10}$ |
| RR 2010 HEU A | $(1.16 \pm 0.02)\times10^{-8}$ | $(1.24 \pm 0.05)\times10^{-8}$ |
| RR 2010 HEU B | $(1.37 \pm 0.02)\times10^{-8}$ | $(1.53 \pm 0.04)\times10^{-8}$ |
| | EK2 | ITU2 |
| NBS U100 | $(6.90 \pm 0.45)\times10^{-10}$ | $(6.81 \pm 0.71)\times10^{-10}$ |

The comparative measurements of some other materials at different locations can be seen in Table 3. The results agree within the "1-sigma" measurement uncertainties, except for "RR 2010 HEU A" and "RR 2010 HEU B". For these two materials we believe the uncertainty is underestimated. Nevertheless, apart from the two outliers, the numbers in Figure 5 and Table 3 suggests that the results may be independent of the detector used for the measurements.

*5.3. Influence of sample age*

If the sample age was not available from destructive measurements (done in-house or taken from the references [20] [21]), then it was either determined by low-background gamma spectrometry [18], [22], [23], or it was estimated from the sample documentation. Often the uncertainty of the sample age is quite large due to various reasons: high uncertainty of the gamma-spectrometric age measurement, vague information available in the sample documentation or the material being a mixture of materials of different ages. Therefore it is important to investigate the influence of the uncertainty of the sample age on the uncertainty of $^{232}U$ measurement.

Figure 6 shows the dependence of the relative uncertainty of the measured $^{232}U/^{238}U$ activity ratio on the uncertainty of the sample age, assuming that all other sources of uncertainty are zero. It can be seen that if the sample is 10 years old and the uncertainty of the age is 10 %, then this contributes about 1 % to the overall uncertainty of the measured $^{232}U/^{238}U$ ratio. If the sample is more than 20 years old, then this contribution is less than 0.5 %, even if the uncertainty of the age is 50%.

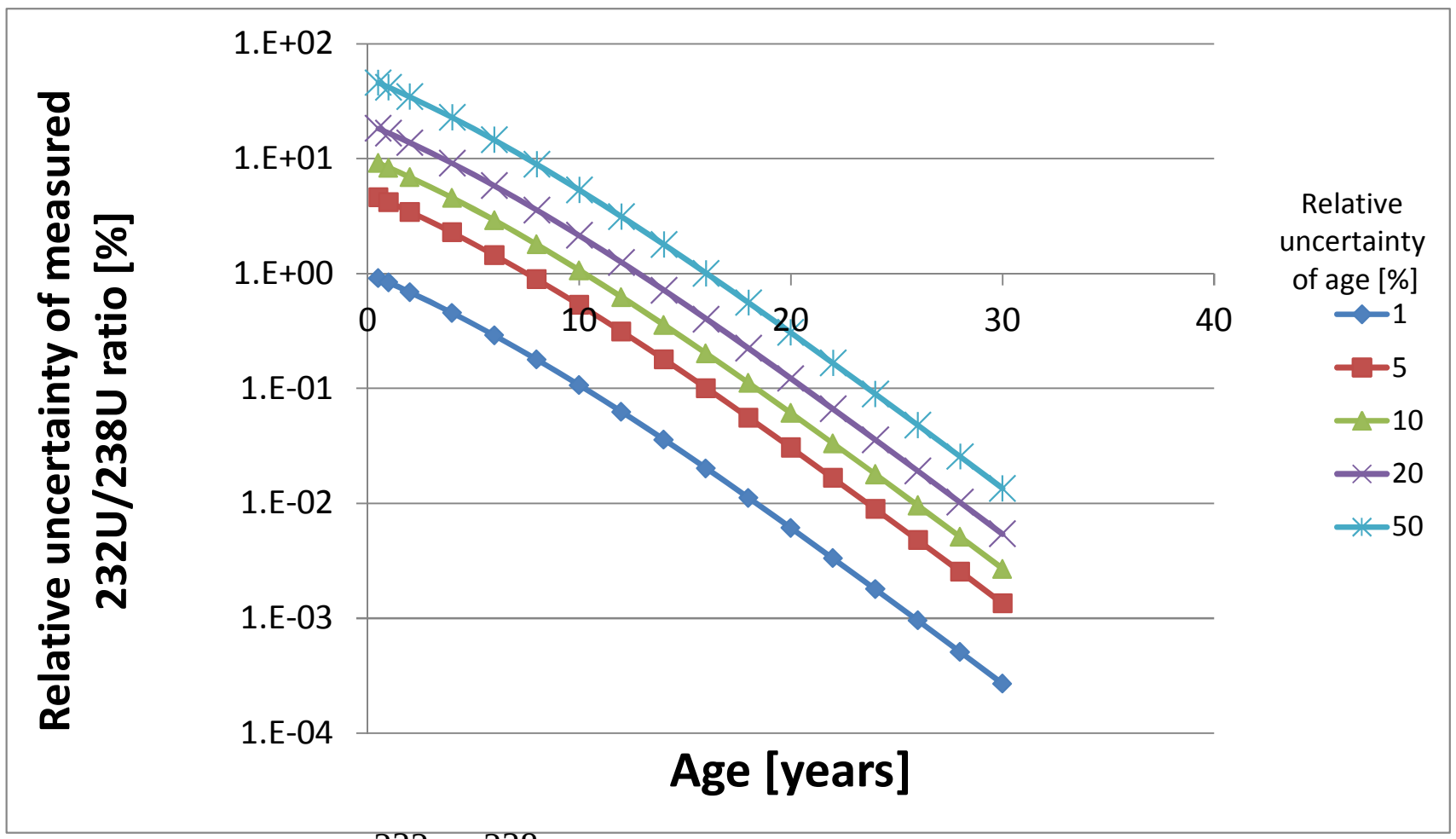

**Figure 6.** Uncertainty of $^{232}U/^{238}U$ mass ratio as a function of age of the sample, for various uncertainties of age. As the sample gets older, the influence of the uncertainty of age becomes less important. This figure was plotted assuming a fixed activity ratio $^{212}Bi/^{238}U$ of 0.03.

The uncertainty of the sample age is much more relevant for extrapolating the $^{232}U$ content back to the time of production of the material, as it can be seen from Figure 7. If the age of the sample is not known then, to be able to calculate the $^{232}U$ content, the age is estimated from the information and assumptions on the history of the material. From Figure 7 we see that this could lead to high uncertainties and bad precision of the $^{232}U$ result. Nevertheless, even this precision is often enough to provide information on the history and origin of the material. For example, as Figure 4 shows, one can easily distinguish between ECGU and uranium which does not satisfy the requirements of ECGU.

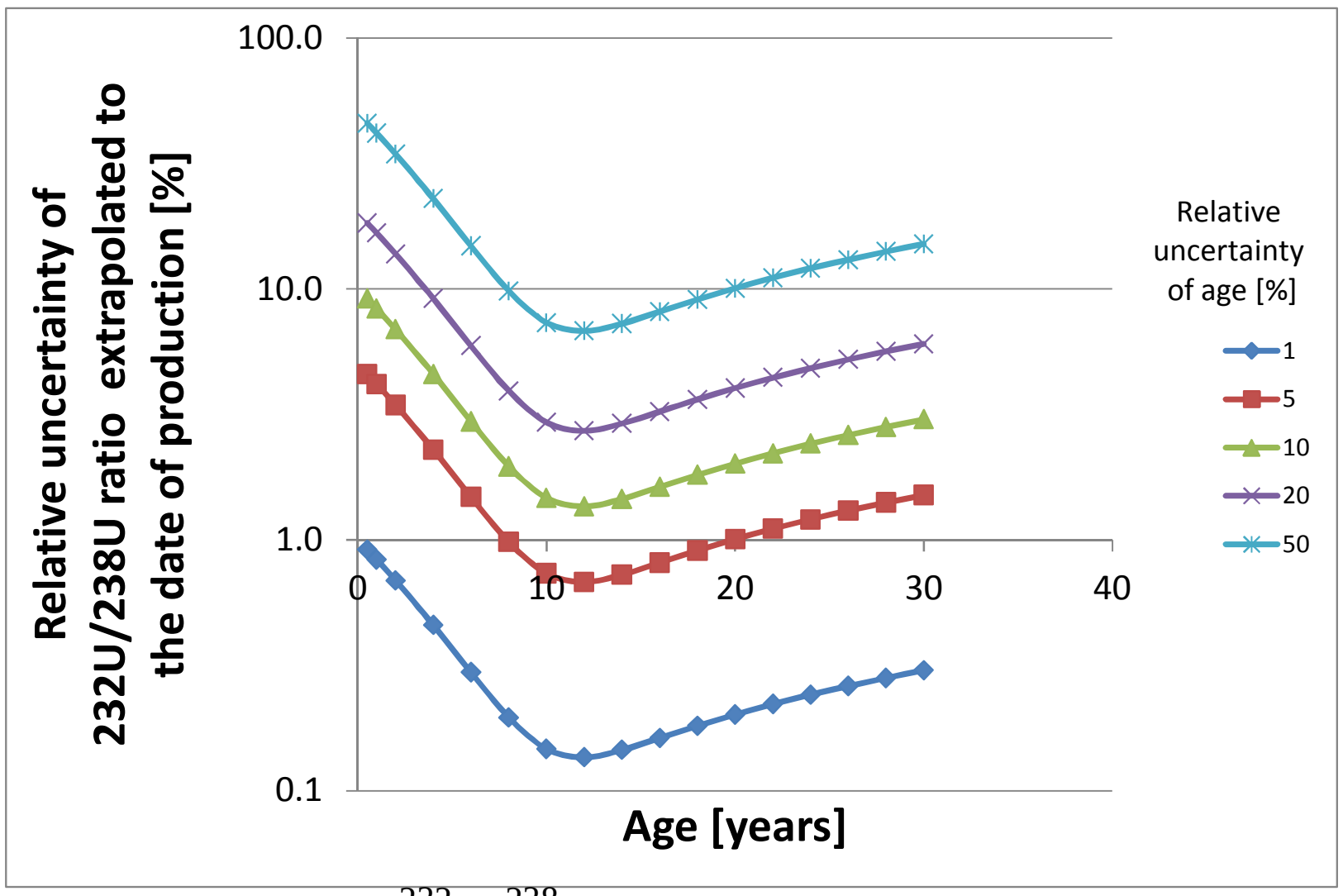

**Figure 7.** Uncertainty of the $^{232}U/^{238}U$ mass ratio extrapolated to the time of production of the material as a function of sample age, for various uncertainties of age. For extrapolating the $^{232}U$ content back to the time of production the uncertainty of age initially decreases with sample age, and becomes again more relevant after about 12 years.

### *5.4. Influence of $^{232}Th$*

$^{232}Th$ is a daughter product of $^{236}U$, which is another indicator of reprocessed uranium together with $^{232}U$. Therefore, if $^{232}U$ is present in the sample, then $^{236}U$, and consequently $^{232}Th$ are also present. In addition, $^{232}Th$ can be also present as an impurity in the sample.

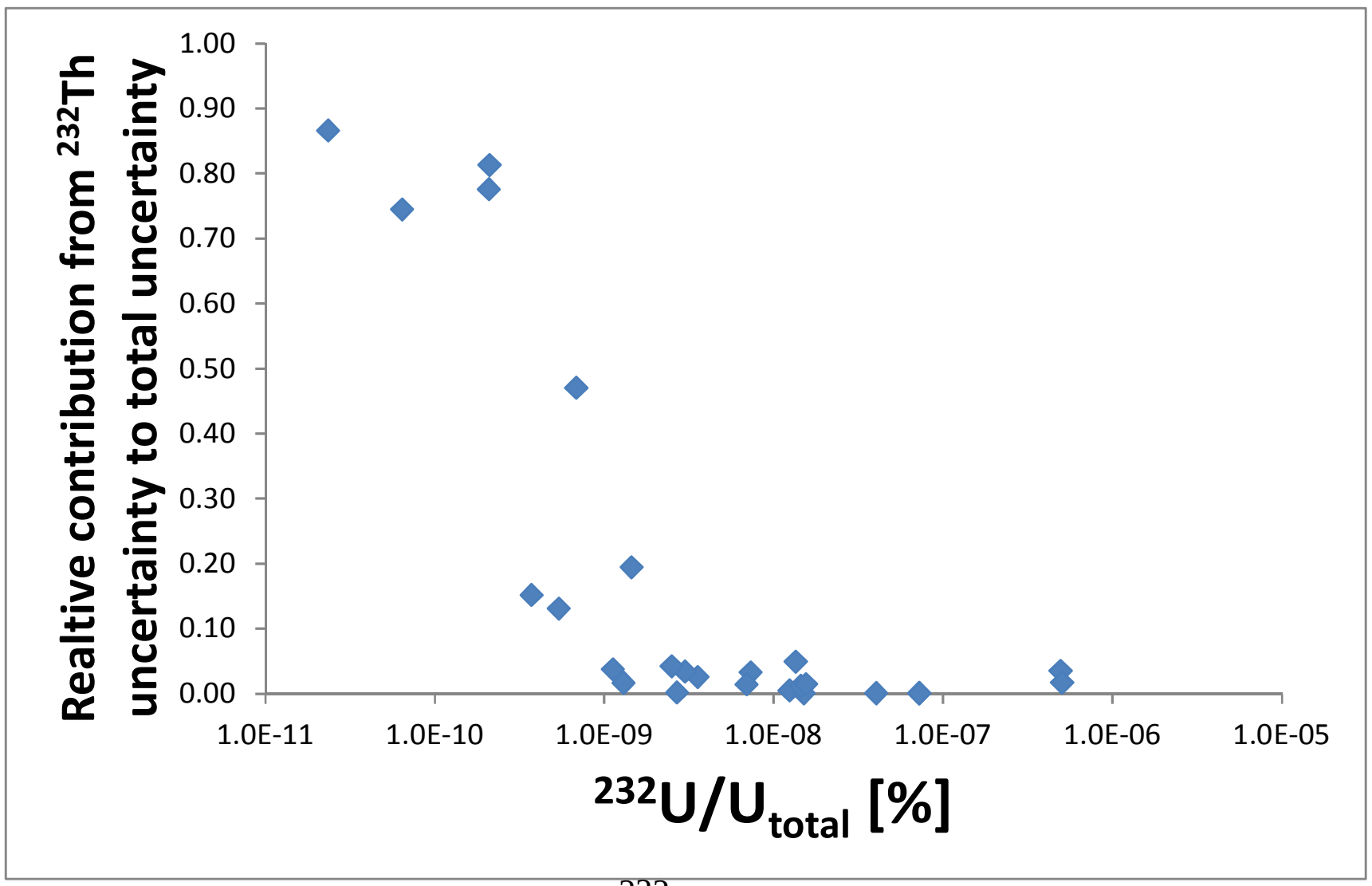

**Figure 8.** Relative contribution from $^{232}Th$ uncertainty to the total uncertainty of the measured $^{232}U$ content, as a function of the $^{232}U$ content. For most measured items the influence of $^{232}Th$ on the uncertainty of $^{232}U$ is negligible, compared to other sources of uncertainty..

We see from the decay chain of $^{232}U$ and $^{232}Th$ (Figure 1) and from equations (1) and (2) that the $^{232}Th$-daugthers present in the samples contribute to the activity of $^{212}Bi$ in the sample. This means that the contribution of $^{232}Th$-daughters has to be subtracted from the measured $^{212}Bi$ activity to accurately determine the $^{232}U$ content. (Note that this contribution is additional to the natural background radiation from $^{232}$Th, which always has to be subtracted before performing the analysis.)

Often the peaks of $^{228}$Ac at 911.316 and 969.171 keV used for calculating the contribution from $^{232}$Th are hidden in the Compton background, so it is not possible to evaluate this term, i.e. it is taken to be zero. Nevertheless, the uncertainty from this zero term still has to be taken into account when calculating the uncertainty of $^{232}$U.

Figure 8 shows the influence of the uncertainty of the $^{232}$Th contribution on the overall measurement uncertainty, as a function of $^{232}$U content. The data in Figure 8 were obtained from the results described in Section 5.1. The relative contribution of $^{232}$Th uncertainty to the overall uncertainty is larger for samples with low $^{232}$U content, but it stays below 1% even for samples with extremely low $^{232}$U content (<$10^{-10}$%). Therefore, the influence of $^{232}$Th on the overall uncertainty is not significant, compared to other sources of uncertainty (e.g. sample age).

### *5.5. Cross-validation with alpha spectrometry*

The $^{232}$U content of a sub-sample from the batch "HU642" has been also measured by alpha-spectrometry [24]. The result given in [24] and the result from gamma spectrometry are compared in Table 4, both extrapolated to the time of production of the material.

**Table 4.** The $^{232}$U/$^{238}$U ratio from alpha spectrometric and gamma spectrometric measurements for the batch "HU642", extrapolated to the time of production of the material.

| | $^{232}$U/$^{238}$U [g/g] | Unc. [g/g] |
|---|---|---|
| Alpha spectrometry | $9.7 \times 10^{-10}$ | $1.2 \times 10^{-10}$ |
| Gamma spectrometry | $7.5 \times 10^{-10}$ | $1.2 \times 10^{-11}$ |
| Difference | $2.2E \times 10^{-10}$ | $1.2 \times 10^{-10}$ |

It can be seen from Table 4 that the difference between the results from the two types of measurement is larger than the combined standard uncertainty of the measurements. However, it still fits within the interval of the expanded uncertainties with a coverage factor of k=2. Note also that both results can identify the sample as "made of reprocessed uranium", versus "contaminated with reprocessed uranium".

An improved alpha-spectrometric method for the determination of $^{232}$U has been presented in [25]. To gain more confidence in the results and uncertainties of both techniques, further comparisons of gamma spectrometry and alpha spectrometry are desirable.

## 6. Discussion

The nuclide $^{232}$U was found in all investigated items, except in the ones made from uranium of natural isotopic composition. For all samples with natural isotopic composition, the $^{232}$U detection limit was less than $10^{-13}$ g/gU, i.e. they all qualify as "Commercial natural uranium" defined in ASTM C 787 [11]. For the rest of the samples, apart from three exceptions, the $^{232}$U/$^{235}$U mass ratio was below < $2x10^{-9}$ g/g $^{235}$U, which is the limit for “Enriched commercial grade uranium” (ECGU) defined in ASTM C 996 [10].

In three items (pellets (HU642 and Romans1 and Romans2) the $^{232}U$ abundance is about 100 times larger than in other items of similar enrichment and it does not satisfy the criteria for ECGU. This may indicate that recycled uranium was deliberately added during their production and they can be considered as "made of reprocessed uranium". For all the rest, $^{232}U$ is probably present only due to cross-contamination with recycled uranium in the enrichment cascade.

Currently we are investigating other mathematical approaches to analyse the data from the gamma spectra. For example, for HEU one can use the peaks of $^{208}Tl$ to construct the relative efficiency curve, instead of the peaks of the $^{238}U$-daughters which are too weak in HEU. For different samples different approaches yield the best results. A detailed comparison of the different approaches is out of the scope of the present work and it will be presented elsewhere.

## 7. Conclusion

Based on the described gamma spectrometric measurements, the samples made of reprocessed uranium can be easily distinguished from samples in which only traces of reprocessed uranium are present. Although the age of the samples influences the measurement of the $^{232}U$ content, the distinction between the two types of samples can be made even if the sample age is not exactly known. The method is fast, non-destructive, does not require any sample preparation and is applicable to a wide range of $^{232}U$ contents, spanning 4 orders of magnitude.

## 8. Acknowledgements

We wish to thank J. Magill and Z. Varga for useful comments on the manuscript.

**Appendix**

**Table 5.** Basic information about the measured items

| Sample ID | Sample type | Measurement date | Age on measurement day [years] | Detector |
|---|---|---|---|---|
| HU590 | Pellet | January 17, 2007 | 17.1 ± 0.3 | EK1 |
| HU597 | Pellet | June 27, 2008 | 15.6 ± 0.5 | EK1 |
| HU598 | Pellet | February 7, 2007 | 13.5 ± 0.5 | EK1 |
| HU642 | Pellets | February 22, 2007 | 13.9 ± 0.2 | EK1 |
| | | September 3, 2008 | 15.4 ± 0.2 | ITU1 |
| HU643 | Pellet | February 2, 2007 | 16.7 ± 0.3 | EK1 |
| | | October 2, 2008 | 18.4 ± 0.3 | ITU1 |
| HU644 | Pellet | February 16, 2007 | 12.9 ± 1.0 | EK1 |
| CBNM 031 | Powder (CRM) | March 12, 2007 | 29.7 ± 1.0 | EK1 |
| | | September 11, 2008 | 31.0 ± 1.0 | ITU1 |
| | | October 22, 2008 | 31.0 ± 1.0 | INR |
| CBNM 071 | Powder (CRM) | March 9, 2007 | 29.7 ± 1.0 | EK1 |
| | | September 10, 2008 | 31.0 ± 1.0 | ITU1 |
| | | October 21, 2008 | 31.0 ± 1.0 | INR |
| CBNM 194 | Powder (CRM) | March 8, 2007 | 29.7 ± 1.0 | EK1 |
| | | September 9, 2008 | 31.0 ± 1.0 | ITU1 |
| | | October 20, 2008 | 31.0 ± 1.0 | INR |
| CBNM 295 | Powder (CRM) | March 8, 2007 | 29.7 ± 1.0 | EK1 |
| | | September 8, 2008 | 31.0 ± 1.0 | ITU1 |
| | | October 16, 2008 | 31.0 ± 1.0 | INR |
| CBNM 446 | Powder (CRM) | March 2, 2007 | 27.7 ± 1.0 | EK1 |
| | | September 4, 2008 | 29.0 ± 1.0 | ITU1 |
| | | October 15, 2008 | 29.0 ± 1.0 | INR |
| NBS-U005 | Powder (CRM) | July 3, 2012 | 54.4 ± 2.0 | ITU1 |
| NBS-U050 | Powder (CRM) | October 12, 2007 | 41.8 ± 4.7 | EK1 |
| NBS-U100 | Powder (CRM) | September 28, 2007 | 38.3 ± 2.0 | EK2 |
| | | November 10, 2006 | 37.0 ± 2.0 | ITU2 |
| NBS-U800 | Powder (CRM) | July 4, 2012 | 53.8 ± 0.5 | ITU1 |
| NBS-U930 | Powder (CRM) | June 27, 2012 | 42.5 ± 5.3 | ITU1 |
| KFKI36 | Powder | December 8, 2003 | 43.0 ± 4.0 | EK1 |
| KFKI90 | Powder | June 14, 2001 | 42.0 ± 3.0 | EK1 |
| RR2001 HEU | Powder | June 19, 2001 | 23.0 ± 3.0 | EK1 |
| RR2010 HEU-A | Metal | April 12, 2010 | 7.04 ± 0.16 | EK1 |
| | | March 23, 2010 | 6.98 ± 0.16 | ITU1 |
| RR2010 HEU- B | Metal | April 4, 2010 | 6.35 ± 0.16 | EK1 |
| | | March 24, 2010 | 6.32 ± 0.15 | ITU1 |
| EK10 | Broken pieces of EK10 fuel pins | July 27, 2006 | 44.2 ± 2.8 | EK2 |
| VVRSM-211 | VVRSM fuel assembly | August 23, 2006 | 27.9 ± 2.2 | EK2 |
| VVRSM-3-051 | VVRSM/3 triple fuel assembly | August 1, 2006 | 7.0 ± 1.0 | EK2 |
| VVRSM-28 | VVRSM fuel | August 22, 2006 | 43.1 ± 2.2 | EK2 |
| | assembly | | | |
| VVRSM-527 | VVRSM fuel assembly | August 24, 2006 | 39.0 ± 2.1 | EK2 |
| KNK15 | Fission chamber | August 3, 2005 | 25.4 ± 5.8 | EK1 |
| Romans1 | Pellet | October 21, 2008 | 1.0 ± 0.2 | ITU1 |
| Romans2 | Pellet | October 24, 2008 | 1.0 ± 0.2 | ITU1 |

**Table 6.** Results of the measurement of the $^{232}U$ content (mass percentages with respect to total uranium). For the materials measured with more than one detector, the value obtained by the EK1 detector (or the EK2 detector, if not measured by EK1) is given. The uncertainties are given with a coverage factor k=1, and they include the counting statistics, the uncertainties of the $^{235}U$ and $^{238}U$ abundance, the uncertainty of the age of the material and the statistical uncertainty of the relative efficiency curve.

| Description / Id | $^{235}U$ / U [%] | Unc. | At the time of measurement $^{232}U/U$ [%] | Unc. | At the time of production $^{232}U$ / U [%] | Unc. | $^{232}U/^{235}U$ [g/g] |
|---|---|---|---|---|---|---|---|
| CBNM031 | 0.3166 | 0.0002 | $1.57 \times 10^{-10}$ | $2.0 \times 10^{-11}$ | $2.12 \times 10^{-10}$ | $2.7 \times 10^{-11}$ | $6.70 \times 10^{-10}$ |
| CBNM071 | 0.7119 | 0.0005 | $<1 \times 10^{-11}$ | | $<1 \times 10^{-11}$ | | |
| CBNM194 | 1.9420 | 0.0014 | $1.8 \times 10^{-11}$ | $1.0 \times 10^{-11}$ | $2.3 \times 10^{-11}$ | $1.4 \times 10^{-11}$ | $1.22 \times 10^{-11}$ |
| CBNM295 | 2.9492 | 0.0021 | $4.79 \times 10^{-11}$ | $7.8 \times 10^{-12}$ | $6.5 \times 10^{-11}$ | $1.1 \times 10^{-11}$ | $2.19 \times 10^{-11}$ |
| CBNM446 | 4.4623 | 0.0032 | $2.84 \times 10^{-10}$ | $1.1 \times 10^{-11}$ | $3.8 \times 10^{-10}$ | $1.6 \times 10^{-11}$ | $8.45 \times 10^{-11}$ |
| NBS U005 | 0.4833 | 0.0005 | $<1 \times 10^{-10}$ | | $<1 \times 10^{-10}$ | | |
| NBS U050 | 4.949 | 0.005 | $9.59 \times 10^{-10}$ | $1.7 \times 10^{-11}$ | $1.460 \times 10^{-09}$ | $7.4 \times 10^{-11}$ | $2.95 \times 10^{-10}$ |
| NBS U100 | 10.075 | 0.010 | $4.70 \times 10^{-10}$ | $2.9 \times 10^{-11}$ | $7.67 \times 10^{-10}$ | $4.7 \times 10^{-11}$ | $7.61 \times 10^{-11}$ |
| NBS U800 | 80.09 | 0.02 | $8.67 \times 10^{-09}$ | $2.2 \times 10^{-10}$ | $1.462 \times 10^{-08}$ | $3.7 \times 10^{-10}$ | $1.83 \times 10^{-10}$ |
| NBS U930 | 93.28 | 0.01 | $2.65 \times 10^{-08}$ | $1.2 \times 10^{-09}$ | $4.07 \times 10^{-08}$ | $2.8 \times 10^{-09}$ | $4.36 \times 10^{-10}$ |
| Single pellet 590 | 0.71121 | 0.00041 | $<1 \times 10^{-10}$ | | $<1 \times 10^{-10}$ | | |
| HU597 | 4.52 | 0.44 | $2.312 \times 10^{-09}$ | $2.7 \times 10^{-11}$ | $2.704 \times 10^{-09}$ | $3.6 \times 10^{-11}$ | $5.98 \times 10^{-10}$ |
| HU598 | 2.04 | 0.02 | $9.92 \times 10^{-10}$ | $3.3 \times 10^{-11}$ | $1.136 \times 10^{-09}$ | $3.8 \times 10^{-11}$ | $5.56 \times 10^{-10}$ |
| HU642 | 2.5121 | 0.0014 | $6.323 \times 10^{-08}$ | $9.6 \times 10^{-10}$ | $7.27 \times 10^{-08}$ | $1.1 \times 10^{-09}$ | $2.89 \times 10^{-08}$ |
| HU643 | 0.25501 | 0.00015 | $1.77 \times 10^{-10}$ | $3.7 \times 10^{-11}$ | $2.10 \times 10^{-10}$ | $4.3 \times 10^{-11}$ | $8.22 \times 10^{-10}$ |
| HU644 | 2.02 | 0.02 | $1.150 \times 10^{-09}$ | $4.2 \times 10^{-11}$ | $1.308 \times 10^{-09}$ | $4.9 \times 10^{-11}$ | $6.49 \times 10^{-10}$ |
| Romans 1 | 4.3207 | 0.0331 | $4.93 \times 10^{-07}$ | $8.3 \times 10^{-08}$ | $4.98 \times 10^{-07}$ | $8.4 \times 10^{-08}$ | $1.15 \times 10^{-07}$ |
| Romans 2 | 4.3207 | 0.0331 | $5.04 \times 10^{-07}$ | $8.5 \times 10^{-08}$ | $5.09 \times 10^{-07}$ | $8.7 \times 10^{-08}$ | $1.18 \times 10^{-07}$ |
| VVRSM No. 211 | 37.29 | 0.41 | $5.26 \times 10^{-09}$ | $1.3 \times 10^{-10}$ | $6.97 \times 10^{-09}$ | $2.3 \times 10^{-10}$ | $1.87 \times 10^{-10}$ |
| VVRSM No. 28 | 36.64 | 0.43 | $1.961 \times 10^{-09}$ | $5.3 \times 10^{-11}$ | $3.03 \times 10^{-09}$ | $1.1 \times 10^{-10}$ | $8.26 \times 10^{-11}$ |
| VVRSM No. 527 | 37.22 | 0.33 | $1.702 \times 10^{-09}$ | $6.1 \times 10^{-11}$ | $2.52 \times 10^{-09}$ | $1.1 \times 10^{-10}$ | $6.77 \times 10^{-11}$ |
| VVRSM/3 No. 051 | 36.72 | 0.24 | $1.459 \times 10^{-08}$ | $5.3 \times 10^{-10}$ | $1.566 \times 10^{-08}$ | $5.9 \times 10^{-10}$ | $4.26 \times 10^{-10}$ |
| EK10 | 10.07 | 0.08 | $3.48 \times 10^{-10}$ | $1.5 \times 10^{-11}$ | $5.43 \times 10^{-10}$ | $2.8 \times 10^{-11}$ | $5.39 \times 10^{-11}$ |
| KNK15 | 90 | 3 | $1.13 \times 10^{-08}$ | $3.8 \times 10^{-09}$ | $1.46 \times 10^{-08}$ | $5.0 \times 10^{-09}$ | $1.62 \times 10^{-10}$ |
| KFKI36 | 36.60 | 0.01 | $2.324 \times 10^{-09}$ | $6.9 \times 10^{-11}$ | $3.58 \times 10^{-09}$ | $1.8 \times 10^{-10}$ | $9.79 \times 10^{-11}$ |
| KFKI 90 | 90.6 | 1.5 | $4.83 \times 10^{-09}$ | $3.8 \times 10^{-10}$ | $7.38 \times 10^{-09}$ | $6.2 \times 10^{-10}$ | $8.14 \times 10^{-11}$ |
| RR 2001 HEU | 89.8 | 0.7 | $5.78 \times 10^{-08}$ | $4.5 \times 10^{-09}$ | $7.29 \times 10^{-08}$ | $6.1 \times 10^{-09}$ | $8.11 \times 10^{-10}$ |
| RR 2010 HEU-A | 92.9030 | 0.0038 | $1.158 \times 10^{-08}$ | $3.9 \times 10^{-10}$ | $1.242 \times 10^{-08}$ | $4.2 \times 10^{-10}$ | $1.34 \times 10^{-10}$ |
| RR 2010 HEU-B | 91.416 | 0.0370 | $1.437 \times 10^{-08}$ | $3.5 \times 10^{-10}$ | $1.531 \times 10^{-08}$ | $3.7 \times 10^{-10}$ | $1.66 \times 10^{-10}$ |